\documentclass{article}
\usepackage{emulateapj}
\usepackage{epsfig}

\newcommand{\Wozniak}{Wo\'zniak}
\newcommand{\Szymanski}{Szyma\'nski}
\newcommand{\Zebrun}{\.Zebru\'n}

\newcommand{\Pietrzynski}{Pietrzy\'nski}
\newcommand{\Soszynski}{Soszy\'nski}

\newcommand{\spc}{\vspace{0.05in}}
\newcommand{\tf}{t_{\rm fwhm}}

\newcommand{\tc}{t_{\rm cross}}
\newcommand{\tC}{t_C}
\newcommand{\tCm}{t_{\langle C\rangle}}
\newcommand{\tint}{t_{\rm int}}
\newcommand{\wbar}{w}
\newcommand{\weff}{w_{\rm eff}}

\begin{document}
\vbox to 0pt{\hfill CWRU-P11-00}
\title{Binary events and extragalactic planets in pixel microlensing}
\author{Edward A. Baltz}
\affil{ISCAP, Columbia Astrophysics Laboratory, 550 W 120th St., Mail Code
5247, New York, NY 10027}
\and
\author{Paolo Gondolo}
\affil{Department of Physics, Case Western Reserve University, 10900 Euclid
Ave., Cleveland, OH 44106-7079
}

\begin{abstract}
Pixel microlensing, i.e.\ gravitational microlensing of unresolved stars, can
be used to explore distant stellar systems, and as a bonus may be able to
detect extragalactic planets.  In these studies, binary-lens events with
multiple high-magnification peaks are crucial.  Considering only those events
which exhibit caustic crossings, we estimate the fraction of binary events in
several example pixel microlensing surveys and compare them to the fraction of
binary events in a classical survey with resolved stars.  We find a
considerable enhancement of the relative rate of binary events in pixel
microlensing surveys, relative to surveys with resolved sources.  We calculate
the rate distribution of binary events with respect to the time between caustic
crossings.  We consider possible surveys of M31 with ground-based telescopes
and of M87 with HST and NGST\@. For the latter, a pixel microlensing survey
taking one image a day may observe of order one dozen binary events per month.
\end{abstract}


\section{Introduction}

Gravitational microlensing has been a powerful probe of massive dark objects in
the galactic halo (Paczy\'{n}ski 1986; Griest 1991; Alcock et al. 1995).  If
the lens is a binary, a pair of caustic crossings may be observed.  The
caustics of the lens system are curves on which the magnification is formally
infinite, and quite large in practice.  In addition to caustic crossing events,
binary lenses can give smaller perturbations to a standard microlensing
lightcurve.  Such perturbations are in principle detectable (Mao \& di Stefano
1995; di Stefano \& Perna 1997).  Binary events yield more information about
the source--lens system than single--lens events, and it is even possible to
image the surface of the source star due to the large magnification.  Caustic
crossing events have been seen towards the galactic bulge and the Small and
Large Magellanic Clouds (Alcock et al. 1997; 2000a; Afonso et al. 1998; Albrow
et al.  1995; 1999; Udalski et al. 2000; Afonso et al. 2000).  Mao \&
Paczy\'nski (1991) studied caustic crossing events in classical microlensing.
They came to the conclusion that approximately 7\% of microlensing events
towards the galactic bulge would exhibit caustic crossings, with the assumption
of perfect time sampling.

The pixel technique extends the feasibility of microlensing experiments by
removing the requirement that individual stars be resolved (Crotts 1992,
Baillon et al. 1993).  The possibility of detecting caustic crossings due to
binary lenses in pixel events is intriguing, as it allows us to study stellar
populations at great distances.  Several groups have successfully used the
pixel, or difference image analysis technique to detect candidate microlensing
events towards M31 (Crotts \& Tomaney 1996; Ansari et al. 1999).  The MACHO
collaboration has also used this technique to identify candidates towards the
LMC, in addition to the classical events (Alcock et al. 1999).  This technique
has a great future as more telescope resources are brought to bear, both on the
ground and in space.  Space based surveys will allow the reach of microlensing
surveys to extend at least as far as the Virgo cluster, at a distance of about
16 Mpc.

In this paper, we extend the Mao \& Paczy\'nski (1991) analysis to pixel
microlensing surveys of stellar systems.  In pixel microlensing, detected
single--lens events tend to have high magnifications, so their rate is
suppressed by a factor of $1/A$, where $A$ is the magnification required for
detection.  However, caustic crossing events always exhibit large
magnifications at the caustic crossings.  This enhances the rate of caustic
crossing events relative to single--lens events in pixel microlensing.  On the
other hand, finite source effects will limit the maximum magnification during a
caustic crossing, rendering some classically detectable caustic crossing events
undetectable in pixel microlensing.  In this paper we show that the larger
region of high magnification is the stronger effect.  As a result, caustic
crossing events in pixel microlensing represent fractions of 10-15\% of
microlensing events, compared with the 2-3\% fraction we find for observable
caustic crossing events with resolved sources.  Finer time sampling would
increase both fractions, and the latter would then agree with the Mao \&
Paczy\'nski (1991) result.

Our primary goal with this work is to show that microlensing events exhibiting
caustic crossings are detectable in pixel microlensing surveys.  To this end,
we estimate the rate of caustic crossing events given some basic assumptions.
Less important to our purposes are considerations of accurately determining the
properties of the source--lens system, which we leave to future work.

\newpage
\section{The two point mass gravitational lens}

We begin with a theoretical discussion of the two point mass gravitational
lens, the appropriate model of a binary star system.  This model has been
studied extensively (Schneider \& Wei\ss\ 1986; Erdl \& Schneider 1993).  We
quote some relevant results.  We take the masses of the lenses to be $M_1$ and
$M_2$, with $M=M_1+M_2$, and we define dimensionless masses
$\mu_{1,2}=M_{1,2}/M$ such that $\mu_1+\mu_2=1$.  Furthermore, we define
$q=M_2/M_1$, with $0\le q\le 1$ without loss of generality.  We take the
standard notation of distances to the source $D_{\rm s}=L$, deflector $D_{\rm
d}=xL$, and deflector--source distance $D_{\rm ds}=(1-x)L$.  We define the
Einstein radius and the Einstein angle for the system as (Einstein 1936)
\begin{equation}
\label{eq:R_E}
R_E=\sqrt{\frac{4GML}{c^2}x(1-x)},\hspace{0.2in}
\theta_E=\frac{R_E}{Lx}.
\end{equation}

We adopt a complex parameterization (Witt 1990) of the lens system. We
introduce complex angular coordinates on the plane of the sky, $z = (\theta_x +
i\theta_y)/\theta_E$.  Given two lenses at angular positions $z_1$ and $z_2$, a
source at $\zeta$ will have images at the solutions $z$ of the lens equation
\begin{equation}
\label{eq:zeta}
\zeta=z-\overline{\alpha(z)}=z-\frac{\mu_1}{\overline{z}-\overline{z}_1}-
\frac{\mu_2}{\overline{z}-\overline{z}_2}.
\end{equation}
Here $\alpha$ is the deflection angle, and the bars indicate complex
conjugation.  We will call the plane described by $\zeta$ the source plane, and
the plane described by $z$ the image plane.  The variables $\zeta$ and $z$ are
angles expressed in units of the Einstein angle $\theta_E$.  A source at
angular position $\zeta$ is at a transverse distance $L \theta_E |\zeta| =
R_E|\zeta|/x$ from the line of sight through the origin of the coordinate
system.  Likewise, a lens at angular position $z$ is at a distance $xL \theta_E
|z| = R_E|z|$ from the line of sight through the origin.  We are interested in
the {\em apparent} physical size of the caustic curves, as this determines the
allowed {\em physical} trajectories of the source which produce caustic
crossings.  The caustic curve, when considered to be in the physical source
plane, is given by $L \theta_E \zeta = R_E\zeta/x$.

In this formalism, the Jacobian of $\zeta(z)$
in eq.~(\ref{eq:zeta}) is
\begin{equation}
J(z)=1- \left| \frac{\partial\alpha}{\partial z} \right|^2 .
\end{equation}
The magnification of an image at $z$ is given by $1/|J(z)|$, with the sign of
$J(z)$ giving the parity of the image. The total magnification is the sum of
the individual magnifications of the images.

The points $z$ where the Jacobian vanishes, $J(z)=0$, trace out the critical
curves.  The inverse images $\zeta(z)$ of the critical curves are the caustic
curves.  When the source is on a caustic, and so its images are on the critical
curve, the magnification is formally infinite.  

Let $d$ be the angular distance between the two lenses in units of the Einstein
angle $\theta_E$. According to the value of $d$, there are three regimes in a
binary lens system: the so--called close, intermediate, and wide binaries.  A
close binary exhibits three caustics, one with four cusps on the line of the
lenses and two with three cusps off the line. The intermediate binary exhibits
one six--cusped caustic, and the wide binary exhibits two four--cusped caustics
on the lens line.  The separation points $d_{\rm CI}$ and $d_{\rm IW}$ between
close and intermediate and between intermediate and wide binaries are as
follows (Erdl \& Schneider 1993; Rhie \& Bennett 1999),
\begin{equation}
d_{\rm IW}=\left(\mu_1^{1/3}+\mu_2^{1/3}\right)^{3/2},\hspace{0.3in}
d_{\rm CI}=\frac{1}{\sqrt{d_{\rm IW}}}\;.
\label{eq:DCIIW}
\end{equation}
In the equal mass case, $\mu_1=\mu_2=1/2$, we have $d_{\rm IW}=2$ and $d_{\rm
  CI}=1/\sqrt{2}$.  As $\mu_1\mu_2\rightarrow 0$, the intermediate binary
vanishes, and the transition between close and wide is at $d_{\rm CI}=d_{\rm
  IW}=1$.

The critical curves are simply parameterized by (Witt 1990)
\begin{equation}
e^{-i\phi}=-\frac{\partial\alpha}{\partial z}.
\label{eq1}
\end{equation}
At each $\phi$, there are four roots of the resulting quartic equation in $z$.
Each root $z_i(\phi)$ ($i=1,...,4$) describes a portion of the critical curve.
Extending the parameter range to $0<\phi<8\pi$, these portions can be patched
together so that they follow each other continuously for every disjoint piece
of the caustic. In this way, we achieve one ``continuous'' parameterization of
the caustic, $z=z(\phi)$.  The continuous parameterization is explicitly given
by
\begin{equation}
z(\phi)=\frac{1}{2}\left[-\tilde{\sigma}+e^{i\phi/2}
\sqrt{e^{-i\phi}\left(-2a-\tilde{\sigma}^2+\frac{2b}{\tilde{\sigma}}\right)}
\right],
\end{equation}
with
\begin{eqnarray}
\tilde{\sigma}&=&e^{i\phi/4}\sqrt{e^{-i\phi/2}\sigma^2}, \\
\sigma &=& \sqrt{-2a/3+u_++u_-},\\
a &=&-\left(d^2\!/2+e^{i\phi}\right),\\
b &=&-d\mu e^{i\phi},\\
\mu &=&\mu_1-\mu_2,\\
u_\pm &=&\sqrt[3]{r\pm\sqrt{s^3+r^2}},\\
s &=&-\left(\frac{d^2-e^{i\phi}}{3}\right)^2,\\
r&=&\left(\frac{d^2-e^{i\phi}}{3}\right)^3+\frac{1}{2}d^2e^{2i\phi}(\mu^2-1).
\end{eqnarray}
Here, $\mu$ gives the degree of asymmetry of the lens, $b$ is something
like a dipole moment about the origin, and the other parameters have no
obvious simple interpretation.  The parameterization of the caustics,
$\zeta=\zeta(\phi)$, follows trivially {}from that of the critical curves,
$z=z(\phi)$, using the lens equation, eq.~(\ref{eq:zeta}).

Lastly, we expand the parameterizations of the critical curves and caustics in
the limits where $d\rightarrow 0,\infty$.  Defining 
$u={\rm exp}(i\phi/2)$ with $0<\phi<4\pi$, we find as $d\ll d_{\rm CI}$:
\begin{eqnarray}
z&\approx&u+\frac{1}{2}\mu d+\frac{3(1-\mu^2)}{8u}d^2,\\
\zeta&\approx&\frac{1}{2}\mu d+\frac{(1-\mu^2)}{8}\frac{3+u^4}{u}d^2,
\end{eqnarray}
with $O(d^3)$ errors and as $d\gg d_{\rm IW}$:
\begin{eqnarray}
z&\approx&\frac{1}{2}d+u\sqrt{\frac{1+\mu}{2}}\left(1+\frac{(1-\mu)}{4}
\frac{u^2}{d^2}\right),\\
\zeta&\approx&\frac{1}{2}d-\frac{1-\mu}{2d}\left(1-\sqrt{\frac{1+\mu}{2}}
\frac{3+u^4}{2ud}\right),
\end{eqnarray}
with $O(d^{-3})$ errors.

\section{Rate of caustic crossing}

In computing the rate of single--lens events, the Einstein angle is used as a
``cross section'' for microlensing.  Any source which passes inside the
Einstein ring is said to be microlensed.  Griest (1991) has shown that the rate
of single--lens events depends linearly on the threshold minimum angular
separation between source and lens, implying that the dependence is on the
angular diameter rather than on the solid angle, which would imply a quadratic
dependence on the impact parameter.  In contrast, the optical depth is a
measure of the probability that a given star is microlensed at a given instant.
This is proportional to the area of the Einstein ring.

To compute the rate of caustic crossing events, we compute the ``angular
diameter'' of the caustic structure in direct analogy to the single--lens case.
We will call it the ``angular width'' of the caustic.  We will appropriately
average this ``angular width'' over the distribution of binary systems.  The
optical depth to caustic crossing would similarly depend on the solid angle
enclosed by the caustic, but the rate of caustic crossing events depends on the
angular width, as we now show.

In a caustic crossing event, the source crosses the caustic an even number of
times, half entering and half exiting the interior of the caustic.  Consider
the rate $\Gamma_0$ at which the sources enter the region limited by the
caustic. The number of caustic crossings per unit time, counting each caustic
peak in the lightcurve, is then $2\Gamma_0$. If the orientation of the caustic
is random, as is the case for binary systems whose separation has no preferred
direction in the sky, we find
\begin{equation}
\label{eq:Gamma}
\Gamma_0 = \Phi \, w,
\end{equation} 
where $\Phi$ is the angle-averaged number flux of the sources, and $w$ is the
``width of the caustic,'' defined in terms of the caustic length $\ell$ by
\begin{equation}
\label{eq:w}
w = \frac{\ell}{\pi}. 
\end{equation} 
We take $w$ and $\ell$ in units of $\theta_E$, and $\Phi$ in the corresponding
units.

Eqs.~(\ref{eq:Gamma}) and~(\ref{eq:w}) are proven as follows. The flux of
sources whose proper motion is in a direction forming an angle $\psi$ with the
binary separation is $\Phi d\psi/(2\pi)$.  Consider an infinitesimal piece of
the caustic $d\ell$, whose interior normal makes an angle $\theta$ with the
fixed axis. The number of sources entering the caustic through $d\ell$ per unit
time is
\begin{equation}
\label{eq:dGamma}
d\Gamma_0 = \int_{\theta-\frac{\pi}{2}}^{\theta+\frac{\pi}{2}}
\frac{d\psi}{2\pi} \Phi \cos(\psi-\theta) d\ell =
 \Phi \, \frac{d\ell}{\pi} = \Phi \, dw.
\end{equation}
The limits of integration restrict the flow of sources to those coming from one
side of the caustic only. Integration of eq.~(\ref{eq:dGamma}) along the
caustic gives eqs.~(\ref{eq:Gamma}) and~(\ref{eq:w}).

Eqs.~(\ref{eq:Gamma}) and~(\ref{eq:w}) apply to any closed curve in the plane.
For example, consider the rate of single lens events. The critical curve of a
single lens is a circle, namely the Einstein ring. Its circumference is its
length $\ell$, and eq.~(\ref{eq:w}) implies that $w$ is the diameter of the
circle. Then eq.~(\ref{eq:Gamma}) states the obvious fact that the number of
sources entering the circle equals the flux times the circle diameter.

This method of computing the cross section of a closed curve counts all entries
to the curve.  A curve that is not convex, such as a caustic curve, can have
multiple entries.  The rate $\Gamma_0$ counts all of these entries. For
example, a binary lens event with 4 caustic crossings is counted twice in
$\Gamma_0$. In other words, the rate $\Gamma_0$ is half the
number of caustic peaks per unit time.

\section{Binary systems}

We now discuss the observed physical parameters of binary systems relevant to
our calculation.  It is well known that a large fraction of stars have
companions, perhaps the majority.  A well--known rule of thumb is that the
distribution of periods $P$ for binary systems is constant in $\ln P$, with
roughly 10$\%$ of binaries in each decade from one--third day to ten million
years.  Less well known is the correlation between the masses of the bound
stars.  At high masses, a correlation is seen, but at low masses, the data are
inconclusive.

We proceed to express the separation of the binary pair in terms of its period
and total mass. With semimajor axis $a$, Kepler's law reads
\begin{equation}
\left(\frac{a}{\rm AU}\right)^3=\left(\frac{M}{M_\odot}\right)
\left(\frac{P}{\rm yr}\right)^2.
\end{equation}
The Einstein radius is
\begin{equation}
\left(\frac{R_E}{\rm AU}\right)=B\sqrt{\frac{M}{M_\odot}},
\end{equation}
with $B=0.28538\sqrt{x(1-x)}10^{D/10}$, and $D$ is the distance modulus to
the source.  Assuming that the projected separation is at maximum, the most
probable situation, we find the relation
\begin{equation}
d=\left(\frac{2}{B}\right)\left(\frac{P}{\rm yr}\right)^{2/3}
\left(\frac{M}{M_\odot}\right)^{-1/6}.
\end{equation}
In the range $P/{\rm yr}=[10^{-3},10^7]$, the distribution in $P$ is simply
$dN/d\log P=1/10$, and is correctly normalized.  Fixing all other quantities,
appropriate for the rate calculation we will do, we thus find the distribution
of binaries in $d$, $dN/d\log d=3/20$.

There is disagreement about the distribution in $q$ of binary systems, ranging
{}from a linear rise $\propto 2.6+2.9q$ (Mazeh et al. 1992) to a power law
decline $\propto q^{-1.3}$ (Patience et al. 1998), with other measurements
falling between (Duquennoy \& Mayor 1991; Trimble 1990).  In the next section
we will find that the effects of the $q$ distribution of binaries are mild.

\section{Size of caustic structure}

We evaluate the mean width of the caustics $\wbar$ for the binary systems
described in the previous section.  In the parametric representation of the
caustic $\zeta=\zeta(\phi)$, its length in units of the Einstein angle is
\begin{equation}
\ell=\int_0^{8\pi}\left|\frac{d\zeta}{d\phi}\right|d\phi=
2\int_0^{8\pi}\left|\mathop{\rm Re}\left(e^{-i\phi/2}\frac{dz}{d\phi}\right)
\right|d\phi.
\label{eq2}
\end{equation}

We have computed the length of the caustic structure for various values of $q$.
The caustic length is not as sensitive to the $q$ distribution as it is to the
lens separation $d$, as we show in figure~1. There we plot the width of the
caustic as a function of $d$, averaging over four distributions in $q$: a
linear rise $\propto 2.6+2.9q$, flat, flat in $\ln q$, and a power law
$q^{-1.3}$ (in the last two distributions we take $q>0.1$). The difference is
only a few per cent. The peaks occur at the average values of the
close--intermediate and intermediate--wide binary separations $d_{\rm CI}$ and
$d_{\rm IW}$.

\vbox{
\noindent
\epsfig{width=3.2in,file=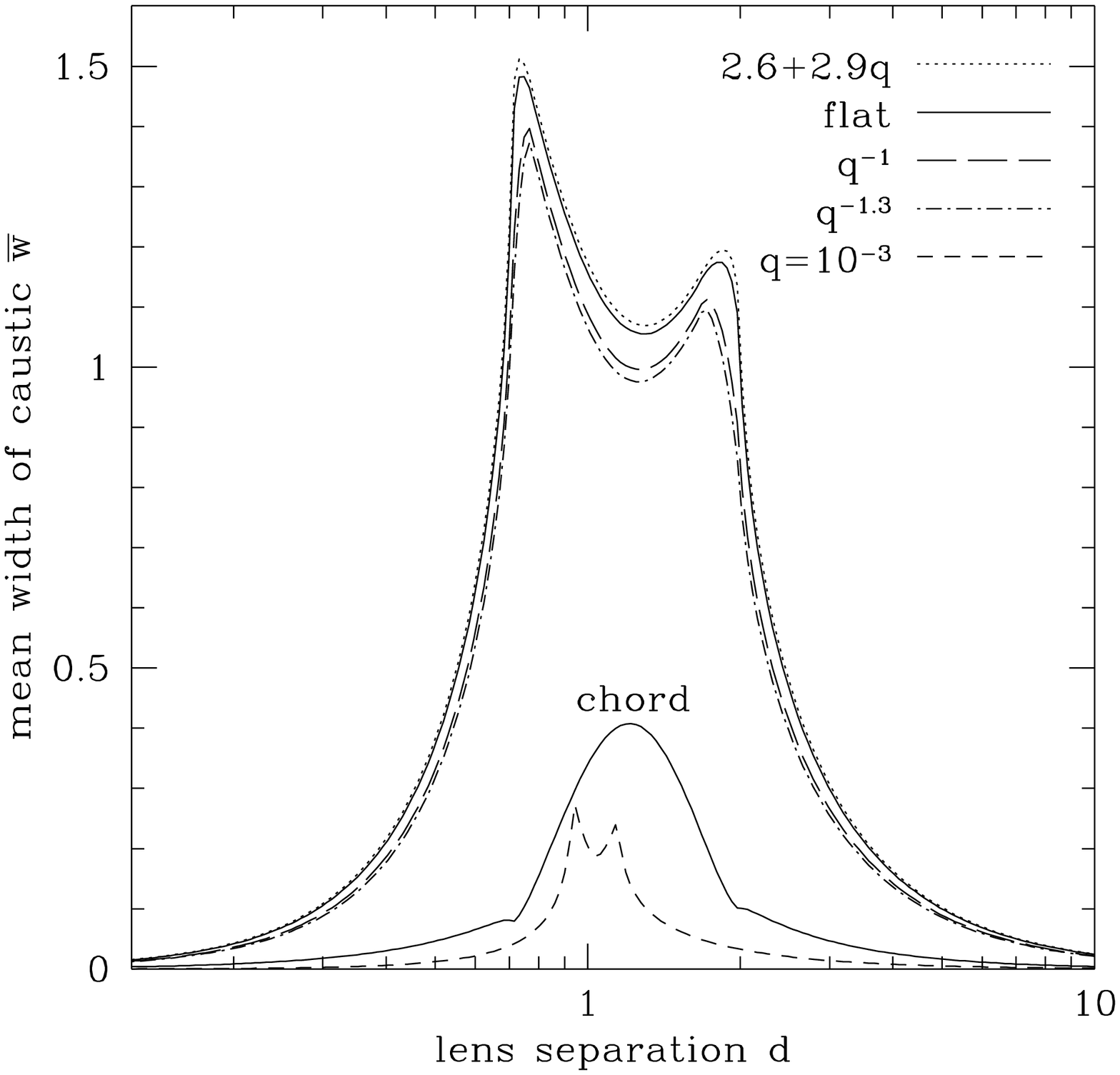}
\hfill\\
{\footnotesize\sc Fig 1.---} Average width $w$ of caustic structure as a
function of lens separation $d$, both in units of the Einstein angle
$\theta_E$.  We have averaged over several distributions in the binary mass
ratio $q$, and also included the case $q=10^{-3}$, of interest for planetary
binaries.  We have also plotted the mean chord length, averaged over the
flat $q$ distribution.
}
\spc

For small and large separations $d$, we have found the leading behavior of the
mean caustic width. In both cases the important caustics are approximately
square.  For $d<d_{\rm CI}$, the linear dimension of the two additional
triangular caustics is suppressed relative to the single, central square
caustic by a factor of $d$.  The mean width is $\wbar=3d^2(1-\mu^2)/\pi$ in the
case $d\ll d_{\rm CI}$, and it is $\wbar=6(1-\mu)\sqrt{2(1+\mu)}/(\pi d^2)$
when $d\gg d_{\rm IW}$.  Thus, we see that the caustic structure is largest
when $d$ is of order unity. Averaging over the four $q$ distributions mentioned
above (namely $2.6+2.9q$, flat, $1/q$, and $q^{-1.3}$, in this order) we find
$\wbar/d^2 = 0.78$, $0.74$, $0.68$, and $0.64$ in the case $d\ll d_{\rm CI}$,
and $\wbar d^2 = 2.0$, $1.85$, $1.63$, and $1.51$ in the case $d\gg d_{\rm
IW}$.

Now that we have computed the cross section for caustic crossing events, we can
compare with that of single--lens events on resolved stars, whose cross section
in Einstein units is 2.  The cross section averaged over angles is $\ell/\pi$,
which needs to be averaged over $d$ and $q$.  We find the ratio of caustic
crossing to single--lens events to be approximately 6.3\%, multiplied by the
fraction of star systems that are binaries.  This is in quite good agreement
with the results of Mao and Paczy\'nski (1991).

In addition to the mean width of the caustic structure, important for
determining the cross section for caustic crossing events, we would like to
know the mean length of chords through the caustic structure.  This quantity is
directly related to the mean time between caustic crossings.  We proceed as
follows.  Let $w(\theta)$ be the projected width of the caustic structure at a
fixed orientation $\theta$, and let $C(\theta,b)$ be the chord length at
orientation $\theta$ and impact parameter $b$.  The area of the caustic
structure is clearly
\begin{equation}
A=\int_0^{w(\theta)}db\,C(\theta,b)=w(\theta)C(\theta),
\end{equation}
where now $C(\theta)$ is the mean chord length at orientation $\theta$.  The
area is of course independent of $\theta$.  We want the mean chord length,
weighted by the width of the caustic.  The mean chord length over all
orientations is now just
\begin{equation}
C=A\int_0^\pi\frac{d\theta}{\pi}\frac{1}{w(\theta)}\frac{w(\theta)}
{\langle w\rangle}=\frac{A}{\langle w\rangle}=\frac{A\pi}{\ell},
\end{equation}
the area divided by the mean cross section.  Note that on the occasions where
there are multiple pairs of caustic crossings, the chord is counted as multiple
separate pieces.

We will need the area and projected diameter of the caustic structure.  In
vector notation, we have $dA=(\vec{r}\times d\vec{r})/2$, which is easily
transformed to the complex parameterization,
\begin{equation}
A=\frac{1}{2}\left|\int_0^{8\pi}{\rm Im}\left(\overline{\zeta}\,
\frac{d\zeta}{d\phi}\right)\,d\phi\right|\hspace{1in}
\end{equation}
\begin{displaymath}
=\left|\int_0^{8\pi}{\rm Im}\left(e^{-i\phi/2}\zeta\right)
{\rm Re}\left(e^{-i\phi/2}\frac{dz}{d\phi}\right)\,d\phi\right|.
\end{displaymath}
The projected diameter along the real axis, at orientation $\theta$, is simply
\begin{equation}
w(\theta)=\frac{1}{2}\int\left|d\,{\rm Re}(e^{i\theta}\zeta)\right|
\hspace{1in}
\label{eq:diameter}
\end{equation}
\begin{displaymath}
=\frac{1}{4}\int_0^{8\pi}\left|e^{i\theta}\frac{d\zeta}{d\phi}+
e^{-i\theta}\frac{d\overline{\zeta}}{d\phi}\right|\,d\phi\hspace{0.5in}
\end{displaymath}
\begin{displaymath}
=\int_0^{8\pi}\left|\cos\left(\theta+\frac{\phi}{2}\right){\rm Re}
\left(e^{-i\phi/2}\frac{dz}{d\phi}\right)\right|\,d\phi.
\end{displaymath}
Note that if this expression is averaged over orientations $\theta$, we recover
the result for the mean width of the caustic structure being proportional to
the length of the caustic.

We have now assembled all of the necessary pieces for calculating the mean
length of chords through the caustic curves.  As for the mean diameter, we
compute this quantity as a function of the lens spacing $d$, averaging over a
distribution in $q$, which we take to be constant in $q$.  The mean chord
length is plotted in figure~1.

We are furthermore interested in the full distribution of chord lengths for a
given lens configuration.  As is done by Han, Park \& Lee (2000), we have
performed a Monte Carlo simulation to find this distribution.  The results are
shown in figure~2, with the exact mean shown as a short dashed line and the
median shown as a long dashed line.  From this simulation, it seems that using
the mean chord length may not be an adequate prescription for designing a
microlensing survey, since the distribution of chord lengths is quite
complicated.  However, we will find that the mean chord length prescription is
usually quite good.

\vbox{
\noindent
\epsfig{width=3.2in,file=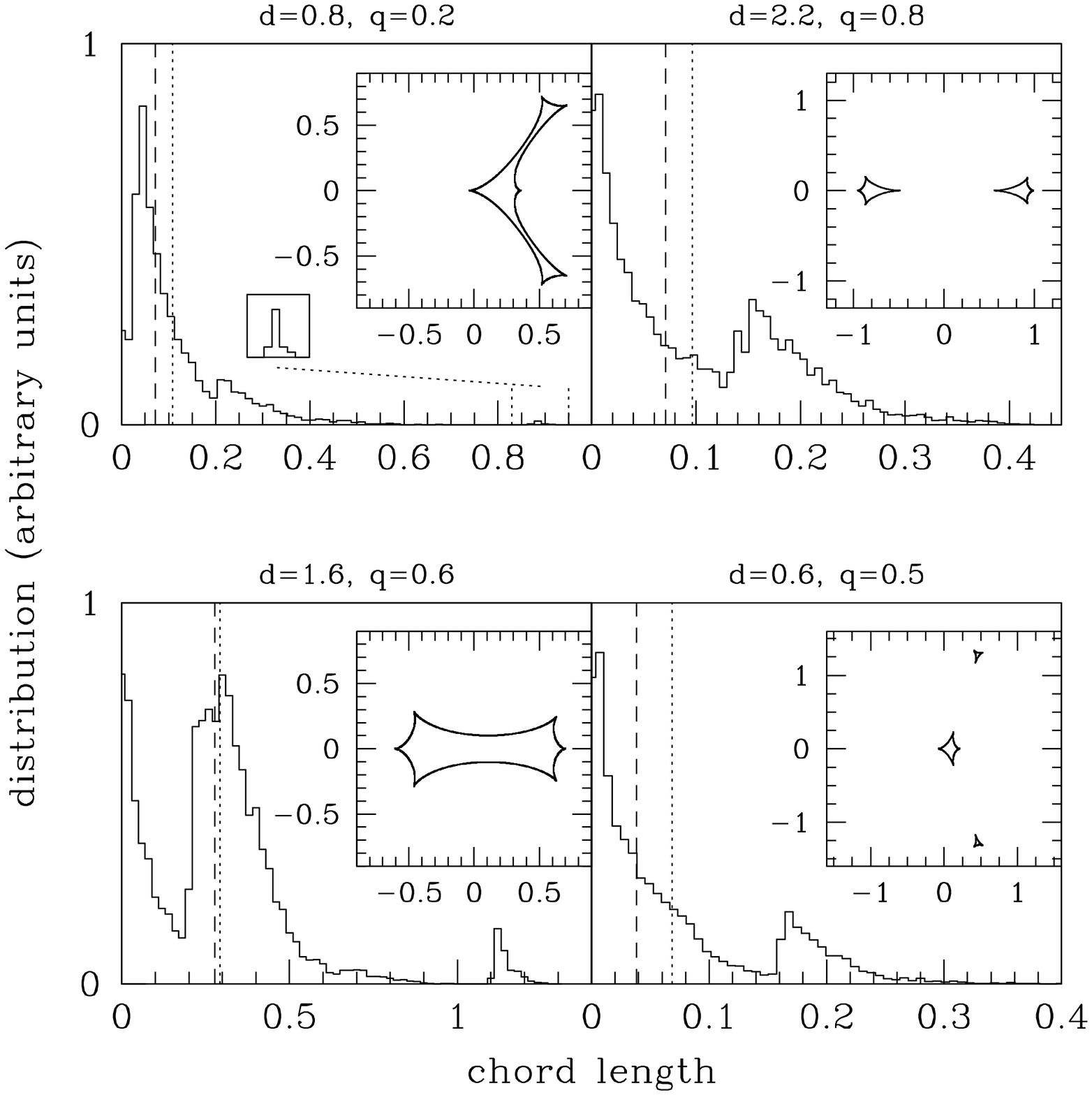}
\hfill\\
{\footnotesize\sc Fig 2.---} Distributions of chord lengths for several choices
of $d$ and $q$.  The exact means of these distributions are indicated by the
short dashed lines, while the medians are shown by the long dashed lines.  The
small panels show the shapes of the caustic curves, with the masses at $x=\pm
d/2$ and the larger mass on the right. The y-axis is in arbitrary units.}
\spc

\section{Lightcurves and Finite size effects}

We illustrate two lightcurves for caustic crossing events in figure~3. Other
examples can be found in the literature cited above. In the upper panels we
show the caustic structure together with the source trajectory in the plane
$(\theta_x,\theta_y)$. We place the lenses along the $\theta_x$-axis at $\pm
d/2$. In the lower panels we plot the lightcurves as functions of
$\theta_x/\theta_E$, which for a uniformly moving source is related linearly to
time.

In the interior neighborhood of a caustic, the magnification diverges as
$K/\sqrt{y}$, where $y$ is the angular distance perpendicular to the caustic,
in units of the Einstein angle (Chang \& Refsdal 1979, Kayser \& Witt 1989).
The flux factor $K$ depends on the lens map through the length of the tangent
vector to the caustic, $T=|T_\zeta|$, as $K=\sqrt{2/T}$.  In the complex
parameterization, the tangent vectors to the critical curve and caustic are
(Witt 1990)
\begin{equation}
T_z= - 2 i \frac{\partial\alpha}{\partial z}
\overline{\frac{\partial^2\alpha}{\partial z^2}},\hspace{0.3in}
T_\zeta=T_z-\overline{\frac{\partial\alpha}{\partial z}}\overline{T}_z.
\end{equation}

Defining the Einstein time as usual $t_E=2R_E/v$, and taking the angle to the
normal as $\delta$, the lightcurve near the caustic crossing at $t=t_0$ is, in
the interior side of the caustic,
\begin{equation}
F=\frac{F_0K}{\sqrt{2\cos\delta}}\sqrt{{\frac{t_E}{\pm(t-t_0)}}},
\label{eq:lightcurve}
\end{equation}
with the $+$ sign ($-$ sign) when entering (exiting) the caustic.  In pixel
lensing $F_0$ is unknown (but see the next paragraph for a possibility to
determine it).  However, there are two caustic crossings, and if the timescales
of the two are compared, the ratio of the geometric factors
$K/\sqrt{\cos\delta}$ at each of the two caustic crossings can be measured.
This may provide some insight into the exact parameters of individual events.

\vbox{
\noindent
\epsfig{width=3.2in,file=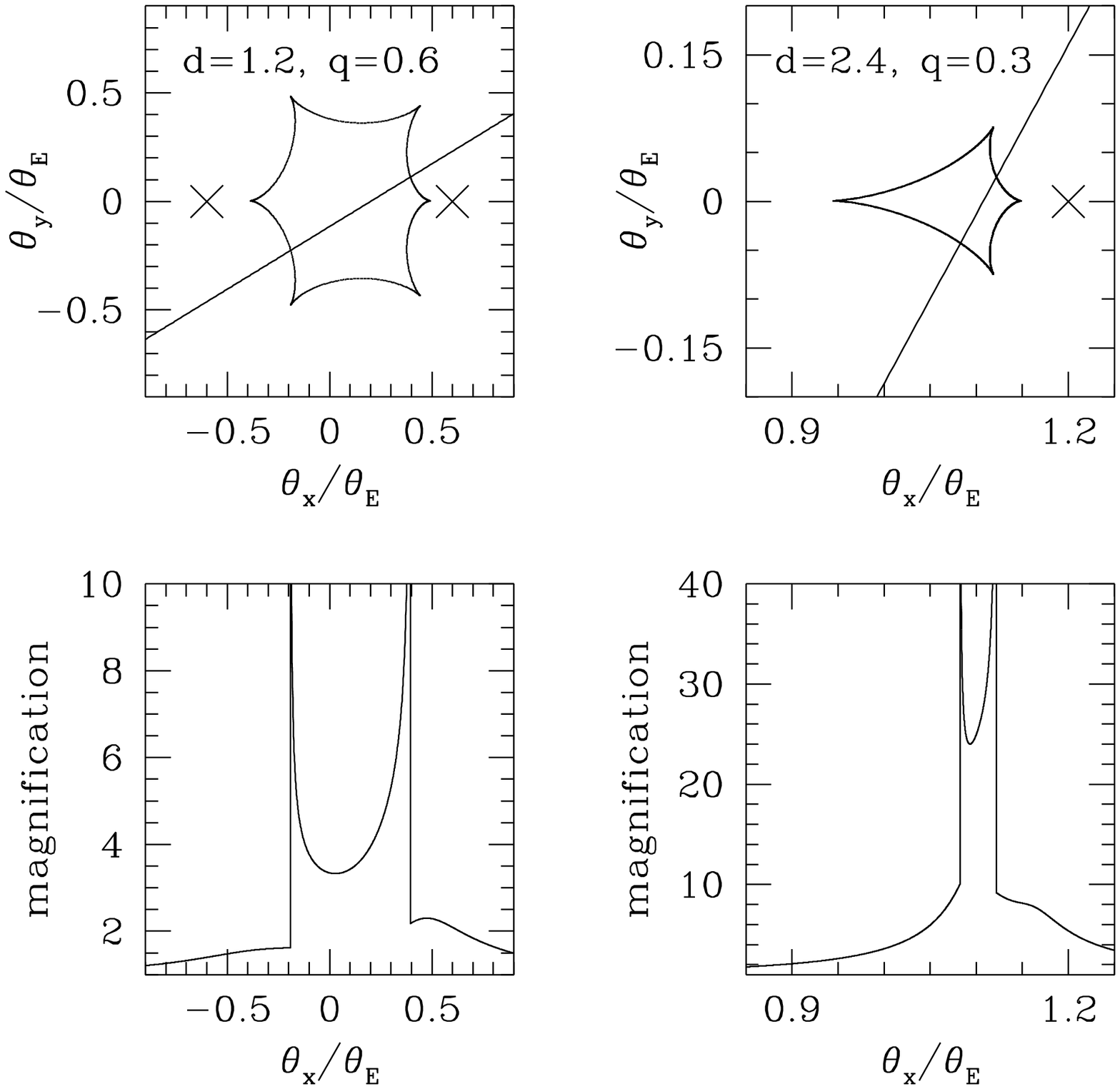}
\hfill\\
{\footnotesize\sc Fig 3.---}
Example source trajectories and lightcurves for caustic crossing events.  The
lenses are marked with crosses, the more massive on the right.
}
\spc

We may gain useful information by measuring the duration of the caustic
crossing.  This is the time over which the disk of the source star intersects
the caustic, and is given simply by
\begin{equation}
\tc=\frac{\theta_s}{\theta_E}\frac{t_E}{\cos\delta}=\frac{2Lx\theta_s}
{v\cos\delta}.
\end{equation}
For star--star lensing in a distant galaxy, $x\approx 1$, $L$ is known, and $v$
and $\cos\delta$ are taken from known distributions.  This can give a handle on
the angular size of the source star.  Furthermore, if the caustic crossing is
seen in multiple wavebands, the color(s) of the source star can be determined.
This is possible because the magnification is the same in all wavebands, thus
comparing the flux {\em increase} in two bands is the same as comparing the
actual (unmeasured) source flux in those bands.  With the color and the angular
size, a rough estimate of the absolute flux $F_0$ can be derived.  It may also
be possible to image the surface of these extragalactic stars as the caustic
passes over them, as has been done towards the SMC (Afonso et al. 1998; Albrow
et al. 1999).  However, this would require intense monitoring to cover the
actual caustic crossing.  It is not the purpose of this paper to discuss this
possibility further.

The magnification during a caustic crossing is critically important to a
calculation of the event rate. While the magnification of a point-like source
is infinite, that of a real source is finite. Since only magnifications higher
than a threshold magnification can be observed, it is crucial to determine the
fraction $f(A\mathord{>}A_0)$ of caustic crossing events that exceed a fixed
magnification $A_0$. The rate of caustic crossings entering the caustic
with magnification $A>A_0$ is then
\begin{equation}
\Gamma = f(A\mathord{>}A_0) \, \Gamma_0 .
\end{equation}
We now determine $f(A\mathord{>}A_0)$.

The maximum magnification of a finite source of angular radius $\theta_s$ is
given by (Schneider \& Wei\ss\ 1987)
\begin{equation}
\label{eq:Amax}
A_{\rm max}=f_sK\sqrt{\frac{\theta_E}{\theta_s}},
\end{equation}
where $f_s$ is a form factor depending on the luminosity profile of the source.
For a uniform disk, $f_s=1.39$, while a limb--darkened profile $\sim
\sqrt{1-r^2/R_s^2}$ gives $f_s=1.47$ (Kayser \& Witt 1989), thus we see that
the profile is relatively unimportant for our purposes. 

{$A_{\rm max}$ in eq.~(\ref{eq:Amax}) is the highest magnification when the
source crosses the caustic. It follows that the fraction of events that cross a
given caustic with magnification $A>A_0$ is equal to the fraction of length of
the caustic in which $K>K_0 \equiv A_0\sqrt{\theta_s/\theta_E}/f_s$,
\begin{equation}
f(A\mathord{>}A_0) = \frac{w({K\mathord{>}K_0})}{w} .
\end{equation}
Taking a distribution of mass ratios constant in $q$, we plot
$w(K\mathord{>}K_0)$ as a function of $d$ for several values of $K_0$ in
figure~4.

\vbox{
\noindent
\epsfig{width=3.2in,file=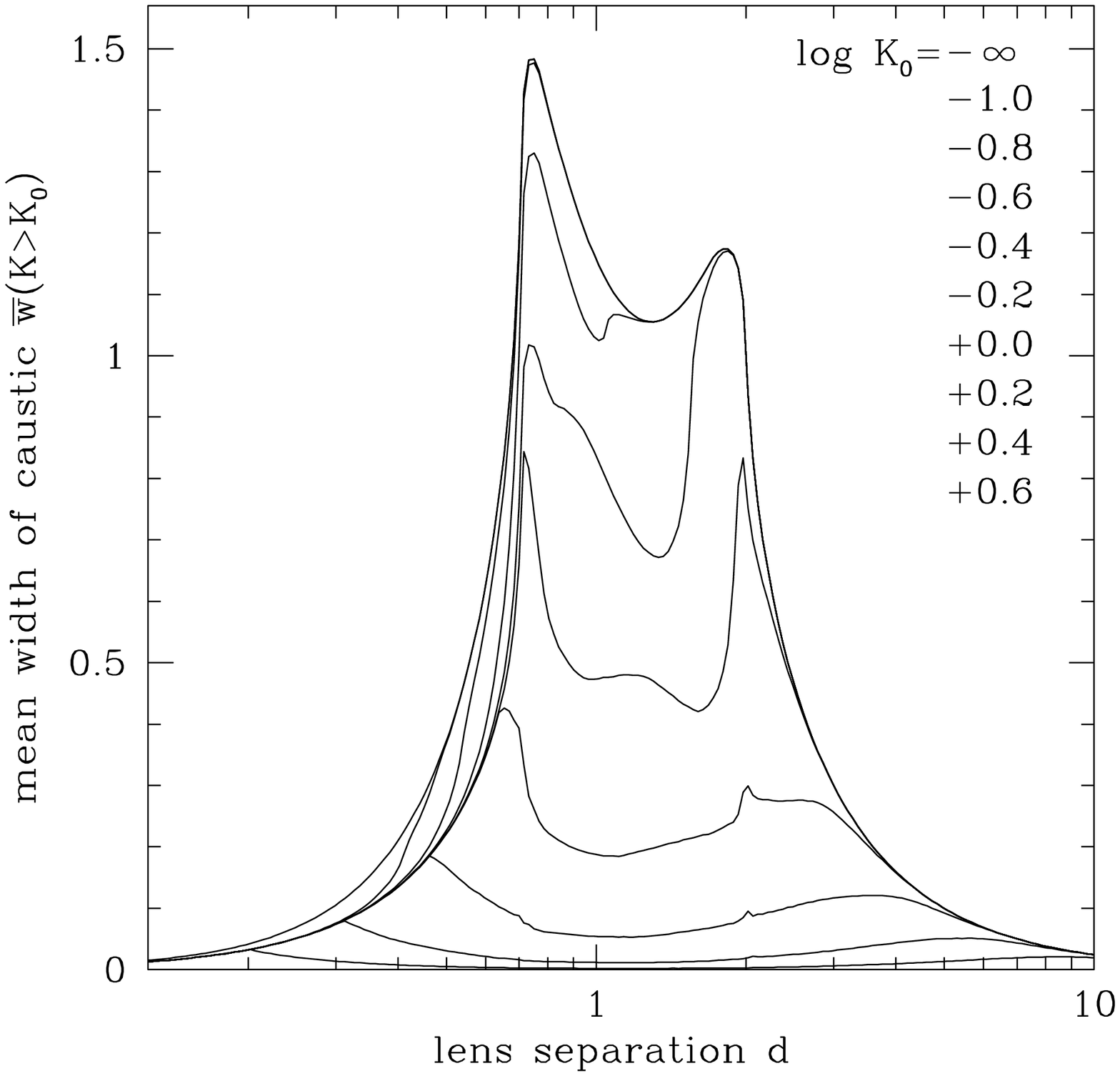}\\
{\footnotesize\sc Fig 4.---}
Effective width of the caustic structure $w(K\mathord{>}K_0)$.  We choose
values equally spaced in $\log K_0$.  The curves decrease in magnitude as $K_0$
increases.
}
\spc

\section{Event rate}

We now compute the rate distribution of caustic crossing events. We let the
velocity distribution of the source and lens population be Maxwellian with
characteristic velocity $v_c$. We denote the transverse velocity of the lens
$v_t$, and the transverse velocity of the line-of-sight $v_l$.  We denote the
total mass of the lens $M$, and the apparent magnitude of the source $m$.  We
define the mass function $\xi(M)=dN/dM$ normalized to unity, and the luminosity
function $\phi(m)$ normalized to the surface brightness.  We denote the
distribution in $d$ as $\Pi(d)=3/(20d\ln 10)$.  The mean width of the caustic
structure in Einstein units is $\wbar$, and we choose a distribution constant
in $q$.

We will compute the rate of caustic crossing events with respect to the time
between caustic crossings $\tC$ in two ways.  First we will use the mean length
of chords to determine this timescale, which is not correct event by event, but
may be acceptable for a rate distribution.  If we define the effective time
between caustic crossings
\begin{equation}
\tCm=\frac{R_E\langle C\rangle}{v_t},
\label{eq:tc}
\end{equation}
where $\langle C\rangle=\langle C(d)\rangle$ is the length of the caustic chord
averaged over the $q$ distribution,} we can write the rate distribution as
\begin{equation}
\frac{d\Gamma}{d\tCm}=2Lv_c^2\int dm\,\phi(m) \int
dM\,\frac{\xi(M)}{M}\int_0^1 dx
\, \rho'(x)
\end{equation}
\begin{displaymath}
\int_{d_{\rm min}}^\infty dd 
\, \Pi(d) 
\, \omega^4
\, e^{-(\omega-\eta)^2}
\, \tilde{I}_0\!\left(2\omega\eta\right)\frac{w(K{>}K_0)}{\langle C(d)\rangle}.
\end{displaymath}
with $d_{\rm min}$ given below, $\omega=R_E(x)\langle C(d)\rangle/(v_c\tCm)$ and
$\eta=v_l/v_c$.  Also, $\tilde{I}_0(x)=I_0(x)e^{-x}$, where $I_0$ is a modified
Bessel function of the second kind.  We note that $\tilde{I}_0(x\gg1)\approx
1/\sqrt{2\pi x}$ and $\tilde{I}_0(0)=1$.

Secondly, we will use the actual caustic crossing timescale, at the cost of
doing an exhaustive Monte Carlo of chord lengths.  Here we define $\tC$ as in
eq.~(\ref{eq:tc}), except that now $C$ is the actual chord length.  The rate
distribution is then
\begin{equation}
\frac{d\Gamma}{d\tC}=2Lv_c^2\int dm\,\phi(m) \int
dM\,\frac{\xi(M)}{M}\int_0^1 dx
\, \rho'(x)
\end{equation}
\begin{displaymath}
\int_0^\infty dv
\, v^3 
\, e^{-(v-\eta)^2}
\, \tilde{I}_0\!\left(2\eta v\right)
\, \weff\left(\frac{v_c\tC}{R_E}\,v,\,K>K_0\right),
\end{displaymath}
with $v=v_t/v_c$, $\weff$ being the effective width as a function of chord
length,
\begin{equation}
\weff\left(C,K_0\right)=\left\langle
w\,\frac{dN}{dC}(K>K_0)\right\rangle_{\theta,q,d},
\end{equation}
$w$ is the projected diameter given in eq.~(\ref{eq:diameter}), and $dN/dC$ is
the probability distribution of chord lengths at a given $\theta,q,d$, computed
by Monte Carlo and normalized to unity at $K_0=0$.  We illustrate the function
$\weff$ for several values of $K_0$ in figure~5.

To fix the minimum flux factor $K_0$, we adopt the convention that the signal
to noise totaled throughout the approximate lightcurve
eq.~(\ref{eq:lightcurve}), be greater than $Q_0=7$.  We assume that samples are
taken daily, (except for the reference case of the Milky Way Bulge, monitored
continuously).  The lightcurve is truncated first by assuming the first sample
is taken at one half day away from the caustic crossing, and then by requiring
that the stellar diameter is less than the length of the chord.  We derive the
signal to noise as follows.  Assume that the telescope collects photons from a
source of magnitude $m$ at a rate $S_010^{-0.4m}$.  An integration of $\tint$
per observation is taken.  The background galaxy light falling on the pixel has
a magnitude $\mu$.  This gives a signal to noise per sample of
$Q=\sqrt{S_0\tint}10^{-0.4(m-\mu/2)}(A-1)$, where we have only considered
photon noise.  Ground based microlensing surveys have achieved noise levels
within a factor of two of the photon noise, and we expect space based surveys
to approach the photon noise limit.  Thus there is a simple relation between
the minimum allowed $Q=Q_0$ and the minimum flux factor $K_0$.

\vbox{
\noindent
\epsfig{width=3.2in,file=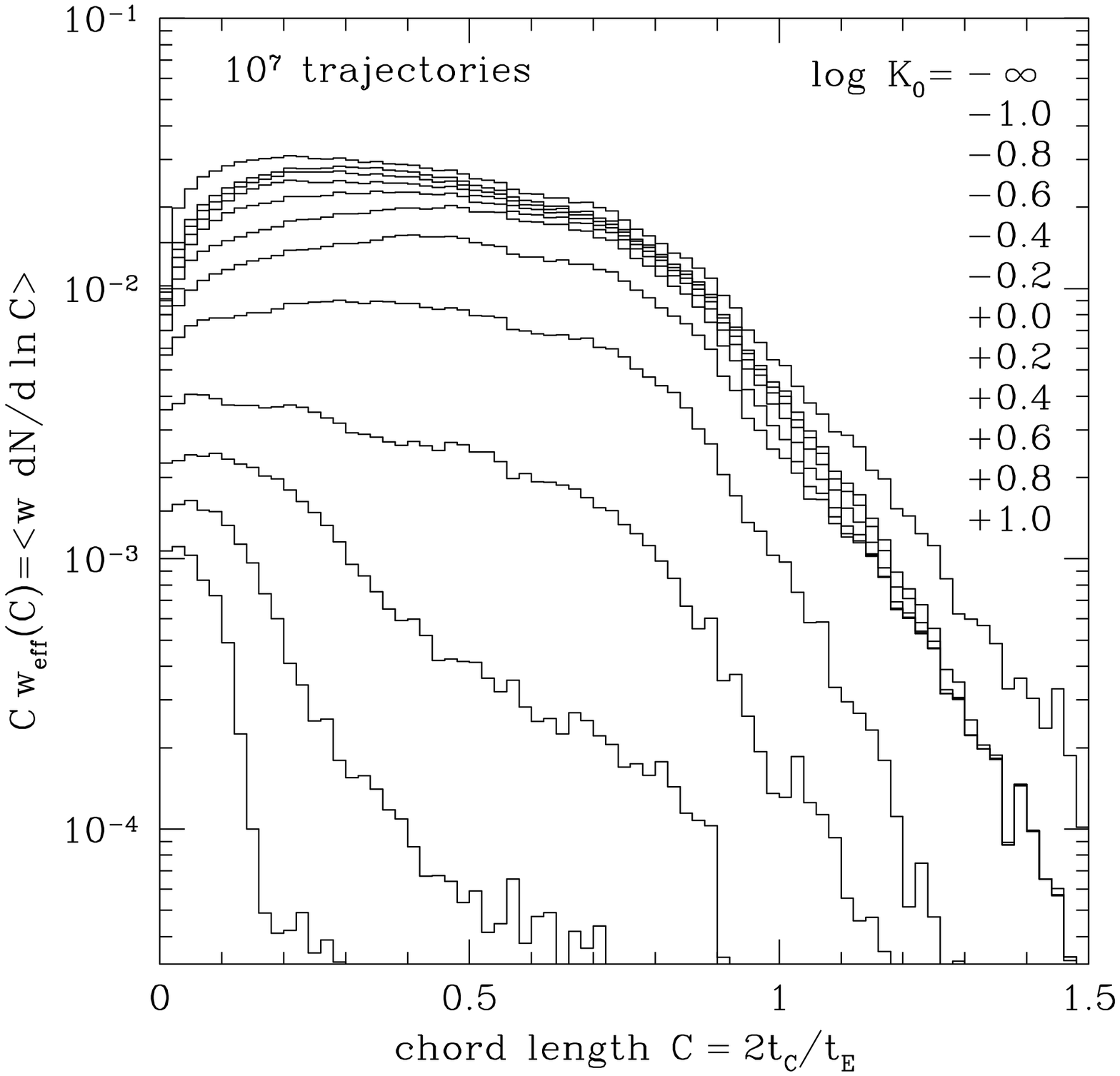} \hfill\\ {\footnotesize\sc Fig 5.---}
Effective width for caustic crossing events.  Ten million trajectories through
caustic structures have been generated for various values of $d$, $q$, and
$\theta$, and the histograms appropriately weighted for several values of
$K_0$.  The $\weff$ histograms have been multiplied by $C$ for clarity.}  \spc

The minimum $d$ allowed is determined by fixing the period at its minimum,
$10^{-3}$ yr, giving $d_{\rm min}=0.02(M/M_\odot)^{-1/6}B(x)^{-1}$.  In
practice, fixing $d_{\rm min}=0.1$ is fine, since $d$ lower than this gives a
very small cross section, and for the mass we are concerned with, it gives a
minimum lens--source distance of 10 pc, which is negligible.  In principle a
maximum $d$ can be determined by the binding of the binary, but it is never
important observationally.

The density $\rho'(x)$ is the effective density of source--lens systems at
separation $x$, and can be determined by integrating the source and lens
positions along the line of sight through the target system.  Assuming a radial
density profile of stars $\rho(r)$ truncated at a radius $R$, we calculate the
column density along a line of sight with impact parameter $b$,
\begin{equation}
N=\int^{\sqrt{R^2-b^2}}_{-\sqrt{R^2-b^2}}\rho\left(\sqrt{b^2+s^2}\right)ds.
\end{equation}
The density along the line of sight is used as a probability distribution for
sources.  Using $y=1-x$, and assuming $L\gg s$ (which furthermore implies that
$R_E$ only depends on $y$ and $L$ and not on $s$), we find
\begin{equation}
\rho'(1-y)=\int^{\sqrt{R^2-b^2}-yL}_{-\sqrt{R^2-b^2}}\frac{1}{N}\,
\rho\left(\sqrt{b^2+s^2}\right)\times
\end{equation}
\begin{displaymath}
\rho\left(\sqrt{b^2+s^2+2Lys+L^2y^2}\right)ds.
\end{displaymath}

Finally, we comment on the rate of binary events with resolved sources.  In
this case there is no maximum $x$, as any caustic crossing event can be
detected.  Thus the integral over the luminosity function $\phi(m)$ is trivial.
The remaining triple integral is simply multiplied by the number of monitored
stars.

\section{Backgrounds and blending}

In any microlensing survey, the issue of backgrounds is crucial.  In
considering binary microlensing events, the lightcurves are basically
characterized by a double spike.  Such events in isolation should be
exceedingly rare.  The most pernicious of backgrounds for single lens events is
the mira variables (Crotts et al. 2000), which have a very long period, but as
they are characterized by a single spike, we are unconcerned with them.
However, if one of the two caustic crossings is missed, the characterization of
the event as microlensing would be very difficult, and only possible with a
very high signal to noise and/or long follow-up period.

\vbox{
\noindent
\epsfig{width=3.2in,file=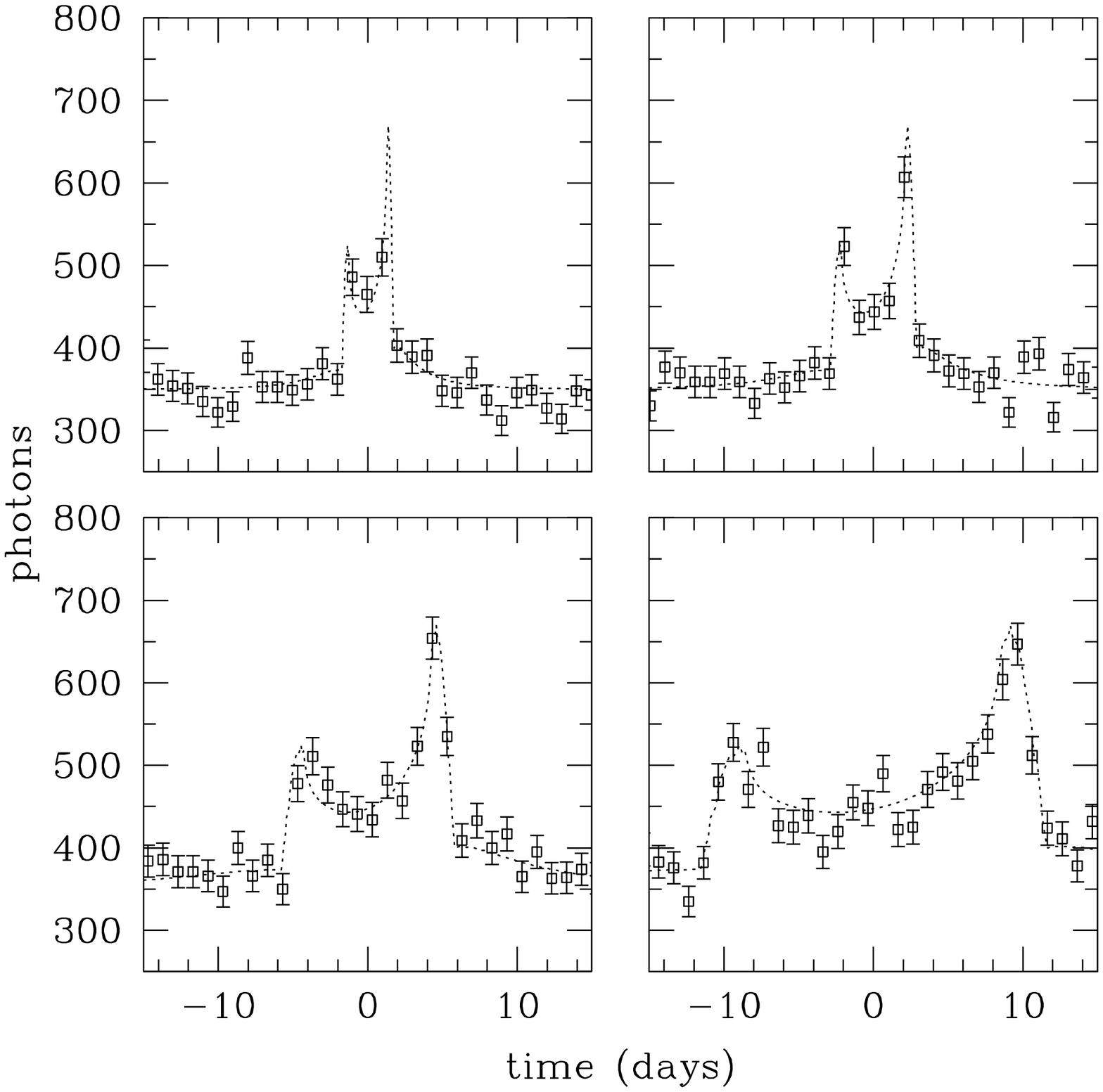} \hfill\\ {\footnotesize\sc Fig 6.---}
Simulated lightcurves in the ACS\@.  The binary system is the same as in the
left panels of figure~3.  We illustrate lightcurves where the time between
caustic crossings is 3, 5, 10, and 20 days.  The brightness is 27.5 mag
pixel$^{-1}$, the source star has an apparent magnitude of 28.5 in $I$, and we
have accounted for the ACS PSF\@.  The dotted curves indicate the theoretical
expectation, including finite source effects.} \spc

In uncovering variability, we must be concerned with the amount of blending of
the lensed star's light.  We will now make a simple estimate of this quantity.
We want to know first the typical magnitude of the lensed star.  Studying a
typical $I$-band luminosity function, such as that of Terndrup, Frogel \&
Whitford (1990), we see that the slope at the highest fluxes is shallower than
0.4, meaning that the most light comes from the brightest stars.  This means
that the majority of detectable microlensing events will be microlensing of
very bright stars.  In fact, typically the rate of such events is expected to
peak somewhere between the surface brightness fluctuation magnitude (which for
the Terndrup et al.\ (1990) luminosity function is roughly
$\overline{M}_I=-1.5$) and the tip of the red giant branch (at
roughly$M_I=-3.5$).  As a rough estimate for M87, if we assume the pixel scale
of the Advanced Camera for Surveys (ACS) on the HST is 0.05\arcsec, and a
surface brightness of roughly 21 mag arcsec$^{-2}$ for most of the field of
view, the pixel brightness is 27.5 mag pixel$^{-1}$, while the surface
brightness fluctuation magnitude is roughly 29.5 for M87, meaning that each
pixel has the light of roughly 6 SBF magnitude stars.  Moreover, we assume a
PSF where roughly 25\% of the light of a star falls on the central pixel.  This
means that if we desire the pixel flux to double, the typical required
magnification (for $M_I=-2.5$) is a factor of 12.  In an 1800 second
integration with the ACS at this surface brightness, roughly 350 photons are
collected, so a $10\sigma$ excess requires only a 50\% magnification of the
pixel flux, or a microlensing magnification that is typically of order six.

In figure~6, we illustrate four representative lightcurves of a binary event in
M87 observed by the HST ACS, sampling daily.  In all cases the geometry is
identical to the left panel of figure~3, but with four different timescales
between caustic crossings of 3, 5, 10, and 20 days.  With a timescale of three
days, though the event is detected at a significant signal to noise, it is not
convincingly a binary.  With a timescale of five days or more, it seems more
likely that binary events can be unambiguous.

Finite source size effects are not severe for the example lightcurves in
figure~6.  The star of magnitude $M_I=-2.5$ has a radius of roughly 65
$R_\odot$ (Gould 1995), and for a solar mass lens system with $D_{ds}\approx5$
kpc, the star subtends roughly one seventh of the angular distance between
caustics along its path.  The dotted curves are the theoretical lightcurves,
including finite source effects for a uniform surface brightness star, computed
according to Gould \& Gaucherel (1997).  According to eq.~(\ref{eq:Amax}), the
limiting counts should be roughly 475 on the left and 600 on the right.  As
shown, the full calculation is not quite that severe.  That the lightcurves
should not be seriously affected is also clear from the fact that the angular
size of the star is small relative to the angular distance between the
caustics.  For the two longer events, it becomes possible to see the actual
caustic crossing happening, though with one image a day the accuracy is poor.
In principle, one could envision an alert system so that more frequent images
are recorded during a suspected caustic crossing.

\section{Example microlensing surveys}

We consider the case of star--star lensing in M87, a giant elliptical in the
Virgo cluster at a distance of 15.8 Mpc.  We take the radial profile of stars
to be
\begin{equation}
\rho(r)=\rho_0\left[1+\left(\frac{r}{\rm kpc}\right)^2\right]^{-\alpha},
\end{equation}
with
\begin{equation}
\alpha=\max\left[1,1+0.275\log\left(\frac{r}{\rm kpc}\right)\right],
\end{equation}
and $\rho_0=3.76$ $M_\odot$ pc$^{-3}$, following Tsai (1993).  We adopt the
$I$-band luminosity function of Terndrup et al. (1990) adjusted to yield a
surface brightness fluctuation magnitude of $\overline{M}_I=-1.5$, and a mass
function $M\xi(M)\propto\exp(-.417\ln M-0.0886\ln^2M)$ with $M$ in solar masses
(Miller \& Scalo 1979).  The velocity dispersion is taken to be $v_c=360$ km
s$^{-1}$.  Typically, stellar radii are $\theta_s\sim 10^{-13}$, while the
typical Einstein radii are of order $\theta_E\sim 2\times 10^{-12}$ for solar
mass lenses, a factor of twenty larger.  We compute the rate of caustic
crossing events observed in surveys using the ACS on the HST, and on a model
NGST\@.  We compare with the rate of single--lens events (Baltz \& Silk 2000),
requiring seven samples detected at $2\sigma$, as Criteria~A of Alcock et
al. (2000b).  We assume that the sensitivity of the ACS is 4.5 times that of
the WFPC2 in the I band, meaning that one photon per second is collected from a
star of magnitude $m_I=25.77$, integrated over the entire PSF.  We model the
NGST as having seven times the efficiency of WFPC2, but with nine times as much
collection area.  These rates are plotted in the upper panels of figure~7.  In
a monitoring program of daily observations of M87 for a period of a month, the
Advanced Camera for Surveys on HST would detect of order one caustic crossing
event, and the NGST could detect of order one dozen caustic crossing events.

Microlensing in M31, the Andromeda Galaxy, has been searched for extensively
using the pixel technique (Ansari et al. 1997, 1999; Melchior et al. 1998,
1999; Crotts and Tomaney 1996; Tomaney and Crotts 1996).  We have computed the
rate of caustic crossing events observed in the bulge of M31, assuming a
CFHT--class telescope and using the bulge model of Kent (1989), with a velocity
dispersion of $v_c=220$ km s$^{-1}$.  The typical star has a radius
$\theta_s\sim 2\times 10^{-12}$, and the typical Einstein radius is
$\theta_E\sim 4\times 10^{-11}$, again a factor of 20 larger.  The fiducial
star has magnitude $m_I=26.88$, giving one photon per second integrated over
the PSF.  Furthermore, we assume that the errors are twice the photon counting
noise, as found by the AGAPE collaboration (Ansari et al. 1997).  We again
compare with the rate of single--lens events, and plot the results in figure~7.
A CFHT survey of the bulge of M31 would detect on the order of one binary event
per month of observations.

Finally we compare the rates of single--lens and caustic crossing events
towards the bulge of the Milky Way.  We take the simple bulge and disk model of
Evans (1994), with no bar.  These rates are plotted in the upper panel of
figure~7.  Note that typically $10^7$ stars are monitored in the galactic
bulge.  We notice that our rates are comparable to the the results of the MACHO
project monitoring of the galactic bulge (Alcock et al. 1997; 2000a).  While we
find that roughly $6\%$ of events should exhibit caustic crossings in this
case, the finite stellar radii remove a significant fraction of these events as
the stellar radius becomes comparable to the distance between the caustics, and
we find that roughly 2-3$\%$ of events will exhibit detectable caustic
crossings.

We find that caustic crossing events typically make up $10\%$, and as much as
$15\%$, of the total rate of stellar events in pixel microlensing surveys,
assuming that all stars are binaries.  This is significantly higher than the
relative rate in microlensing with resolved sources, typically 2-3$\%$.  Thus,
the larger effective cross section is more important than the more severe
finite source effects.

These estimates all assume only self--lensing in the target galaxy.  Of course
there is the possibility that lenses in the halo of the Milky Way could
contribute.  Unless there is a significant population of halo lenses, the rate
for stellar lenses in the Milky Way would be substantially smaller than the
self--lensing rate in the target galaxy, since the scale height of the disk is
much smaller than the size of the spheroids of the target galaxies.

\vbox{
\noindent
\epsfig{width=3.2in,file=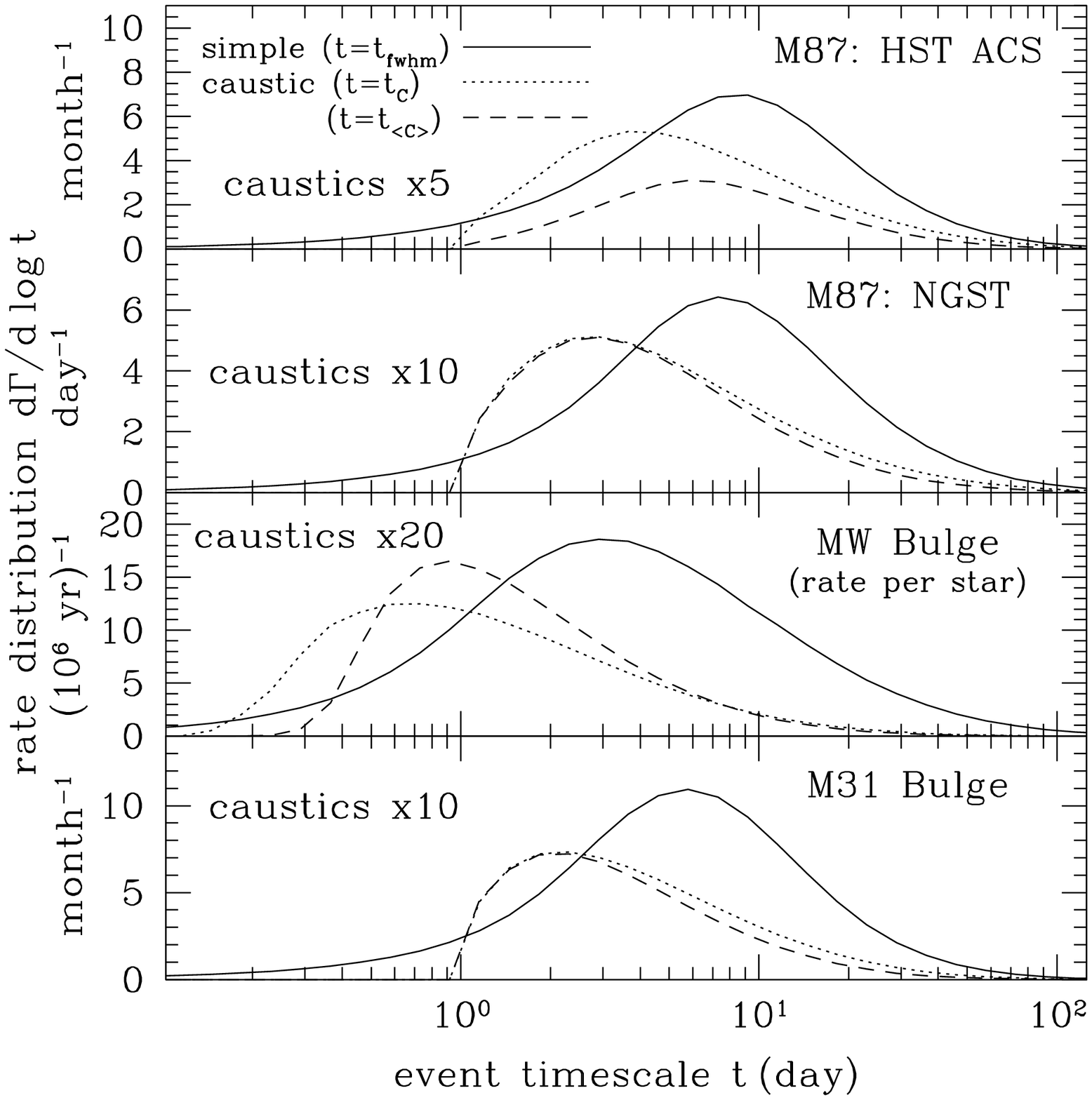} \hfill\\ {\footnotesize\sc Fig 7.---} Rate
of microlensing events.  In all cases the solid lines indicate the rate of
single--lens events as a function of the full--width at half maximum timescale
$\tf$, the dotted lines indicate the rate of caustic crossing events as a
function of crossing time $\tC$, and the dashed lines indicate the caustic
crossing rate using the mean crossing time $\tCm$.  {\em Top panel:} M87 events
using the Advanced Camera for Surveys on HST\@.  {\em Second panel:} M87 events
using the NGST\@.  {\em Third panel:} Milky Way bulge events for resolved
stars.  This rate is quoted per monitored star.  {\em Bottom panel:} Events in
the bulge of M31.  We use the improved peak threshold trigger for single--lens
events in M31 and M87.  A Miller--Scalo mass function is assumed in all
systems.}  \spc

{}From figure~7 we see that the time between caustic crossings is typically of
order several days.  Sampling daily would allow a sizeable fraction of the
lightcurves to be well--measured.  We also see that for long timescale events,
the timescale $\tCm$ gives a very good approximation to the true rate (using
$\tC$), but for the shorter events, it is less accurate.  We notice that the
agreement between the two is excellent for the NGST and for the M31 bulge.  In
these cases, the required magnification is quite modest, of order a factor of
two for a star with the surface brightness fluctuation magnitude.  The ACS
requires significantly higher magnifications, spoiling the excellent agreement
between the rate with respect to the two timescales $\tC$ and $\tCm$.  For the
Milky Way Bulge, the agreement between the two timescales is quite good, though
the (quite broad) peaks disagree by a factor of about two in timescale.  We
thus find that the timescale $\tCm$ provides a surprisingly good estimate of
the rate of events with respect to the time between caustic crossings.

Finally, we comment that with daily sampling, the fine details of the caustic
crossing will not be well covered.  If we choose a finer sampling scheme, with
the same telescope resources, the expected number of events will be much
smaller.  We feel that it is important first to identify binary lensing events.
If this is shown to be possible, then more sophisticated sampling strategies
should of course be considered.

\section{Extragalactic planets}

As an aside, we apply the previous analysis to a pixel microlensing search of
extragalactic planets.

Discovering planets outside our own solar system has been an important
scientific goal for many years.  The planets that have been discovered so far
were all found by detecting the wobble of the star that they orbit.  The first
detection was of a planet approximately the mass of Earth, in orbit around a
pulsar (Wolszczan \& Frail 1992; Wolszczan 1994).  More recently, planets have
been detected around main sequence stars (see Marcy, Cochran \& Mayor 2000 for
a review).  We would like to extend the range of planetary searches to distant
galaxies.  Gott (1981) was the first to point out that gravitational lensing
might be used to detect low--mass objects in distant galaxies.  Here we make
another suggestion: pixel microlensing surveys may be able to detect
Jupiter--mass planets as far away as the Virgo cluster, by capitalizing on the
unique nature of binary microlensing events.

Previous work has shown that planets might be detected in microlensing events
in the bulge of the Milky Way galaxy, or in the Small and Large Magellanic
Clouds, which are small galaxies in orbit around the Milky Way (Mao \&
Paczy\'{n}ski 1991, Gould A. \& Loeb, A. 1992, Griest, K. \& Safizadeh,
N. 1998).  Stars in the bulge and in the Magellanic clouds can be resolved
easily, and surveys routinely monitor of order ten million stars for
microlensing events.  Evidence for a planet orbiting a binary star system in
the Milky Way bulge has recently been claimed in a joint publication of the MPS
and GMAN collaborations (Bennett et al. 1999), although the data can also be
interpreted as a rotating binary system without a planet (Albrow et al. 2000).

A solar-type planetary system can be described to first approximation as a
binary object, consisting primarily (in mass) of the Sun and the planet
Jupiter. In such systems $q$ is very small, for example $q\approx10^{-3}$ for
the Sun and Jupiter.  In figure~8, we show two example lightcurves of
microlensing events, together with the trajectories of the source stars
relative to the caustic curves.  In both cases the star has a companion one
one-thousandth as massive, like the Sun--Jupiter system.

We have calculated the rate of planetary events observing M31 with a telescope
like the Canada-France-Hawaii telescope (CFHT) on Mauna Kea.  We assume that
every star has a companion that is one one-thousandth as massive, just like the
Sun and Jupiter.  Furthermore, we assume, as is true for known binary stars,
that the distribution of orbital periods is such that ten percent of such
systems lie in each decade of period, from a third of a day to ten million
years.  This gives a 10\% probability that a star has a companion between one
and five AU, in rough agreement with observational findings (Marcy, Cochran \&
Mayor 2000).  With a long--term monitoring program observing four times daily
(using the necessary global network of telescopes), and relaxing the threshold
to be three sigma in the two images appearing inside the caustic, we expect
about one caustic crossing planetary event every two years.  This neglects
other types of microlensing anomalies, such as those discussed by Covone, de
Ritis \& Marino (1999).

To increase the chances to detect planetary systems in distant galaxies, we
require a space telescope such as the proposed Next Generation Space Telescope,
to be launched around 2009.  This will be a large (about 7 meters in diameter)
infrared telescope at a Lagrange point of the Earth--Sun system, and it will
be more than ten times as sensitive as the Hubble Space Telescope.

\vbox{
\noindent
\epsfig{width=3.2in,file=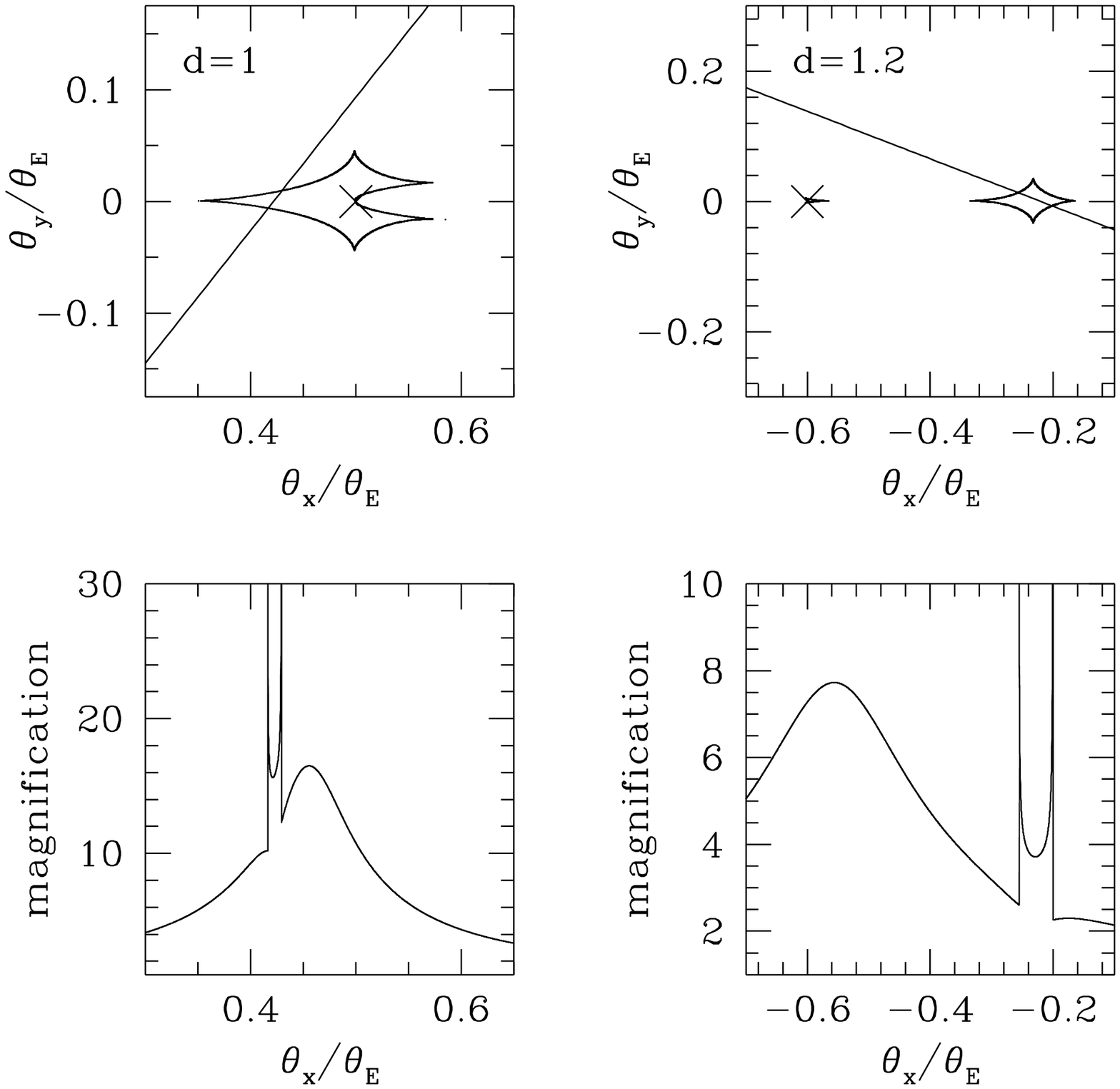} \hfill\\ 
{\footnotesize\sc Fig 8.---}
Microlensing events due to a planetary system.  We illustrate the
trajectory of a source star over the caustic curves of a lens star and its
planet.  We then illustrate the observed magnification of the source star along
its trajectory.  The cross indicates the star's position, whereas the planet
lies off the plots at $-0.5$ and $0.6$, respectively.  The plots are in units
of the Einstein angle $\theta_E$, which is the characteristic angular scale of
a microlensing event.  Also shown is the star--planet separation $d$ in units
of $\theta_E$.}
\spc

We have calculated the rate of events we might detect with the NGST observing
the giant elliptical galaxy M87 in the Virgo cluster, at a distance of 15.8
Mpc.  With the same assumptions as the M31 calculation, we find that an NGST
survey of three month's duration, taking four images each day, should be able
to detect of order four planetary systems.  We find that such a survey is most
sensitive to events where the separation between caustic crossings is about
five days. An alert system for microlensing events would allow more frequent
measurements of the light curve during the caustic crossings, with the
possibility of determining the orbital parameters of the planetary system.

Our results for the rate of caustic crossing planetary events indicate that the
rates are probably too small for this technique to be a feasible method of
detecting extragalactic planets.  We have made overly generous assumptions, and
still the detection probabilities are marginal.

\section{Conclusions}

We have found that caustic crossing binary events should typically make up
$10\%$, and as much as $15\%$, of the total rate of events in pixel
microlensing surveys.  This is significantly higher than the relative rate in
microlensing with resolved stars, which is typically 2-3$\%$. The enhancement
is due to the larger magnification of caustic crossing events with respect to
single-lens events. The suppression in the count rate expected from
finite-source effects is not large enough to win over the enhancement from
larger regions of high magnification.

In examples of pixel lensing surveys, we find that on the order of one binary
event per month of observation should be observable in a CFHT survey of the
bulge of M31. Also of the order of one caustic crossing event per month of
daily observations of M87 should be detectable with the Advanced Camera for
Surveys on HST\@. The latter rate would increase to of order one dozen caustic
crossing events per month with the NGST\@.  Searches for extragalactic planets
with the microlensing technique discussed in this paper do not seem
particularly promising.

\acknowledgments

We would like to thank the referee for constructive comments that have added
much to the original manuscript, and for making us aware of a simplification of
eq.~(\ref{eq:DCIIW}).  E.\ B.\ thanks P.\ Podsiadlowski for productive
conversations concerning binary star systems.  We thank J.\ Silk for useful
comments and hospitality at Oxford University, where this research was begun.
E.\ B.\ was supported in part by grants from NASA and DOE.

\end{document}